\documentclass[superscriptaddress,nofootinbib ]{revtex4-2}

\usepackage{natbib}
\usepackage{hyperref}
\usepackage{microtype}
\usepackage[utf8]{inputenc}
\usepackage{amsmath}
\usepackage{tabularx}
\usepackage[bottom=3cm, top=2.7cm, left=2.7cm, right=2.7cm]{geometry}
\usepackage[dvipsnames]{xcolor}
\usepackage{subcaption}
\usepackage{graphicx} 
\usepackage{units}
\usepackage{upgreek}

\graphicspath{ {data} }

\begin{document}
\title{First measurements of remoTES cryogenic calorimeters: easy-to-fabricate particle detectors for a wide choice of target materials}

\author{G.~Angloher}
\affiliation{Max-Planck-Institut f\"ur Physik, 80805 M\"unchen - Germany}

\author{M.R.~Bharadwaj}
\affiliation{Max-Planck-Institut f\"ur Physik, 80805 M\"unchen - Germany}

\author{I.~Dafinei}
\affiliation{INFN - Sezione di Roma, 00185 Roma - Italy}

\author{N.~Di~Marco}
\affiliation{Gran Sasso Science Institute, 67100 L'Aquila - Italy}
\affiliation{INFN - Laboratori Nazionali del Gran Sasso, 67010 Assergi - Italy}

\author{L.~Einfalt}
\affiliation{Institut f\"ur Hochenergiephysik der \"Osterreichischen Akademie der Wissenschaften, 1050 Wien - Austria}
\affiliation{Atominstitut, Technische Universit\"at Wien, 1020 Wien - Austria}

\author{F.~Ferroni}
\affiliation{Gran Sasso Science Institute, 67100 L'Aquila - Italy}
\affiliation{INFN - Sezione di Roma, 00185 Roma - Italy}

\author{S.~Fichtinger}
\affiliation{Institut f\"ur Hochenergiephysik der \"Osterreichischen Akademie der Wissenschaften, 1050 Wien - Austria}

\author{A.~Filipponi}
\affiliation{Dipartimento di Scienze Fisiche e Chimiche, Universit\`a degli Studi dell'Aquila, 67100 L'Aquila - Italy}
\affiliation{INFN - Laboratori Nazionali del Gran Sasso, 67010 Assergi - Italy}

\author{T.~Frank}
\affiliation{Max-Planck-Institut f\"ur Physik, 80805 M\"unchen - Germany}

\author{M.~Friedl}
\affiliation{Institut f\"ur Hochenergiephysik der \"Osterreichischen Akademie der Wissenschaften, 1050 Wien - Austria}

\author{A.~Fuss}
\affiliation{Institut f\"ur Hochenergiephysik der \"Osterreichischen Akademie der Wissenschaften, 1050 Wien - Austria}
\affiliation{Atominstitut, Technische Universit\"at Wien, 1020 Wien - Austria}

\author{Z.~Ge}
\affiliation{SICCAS - Shanghai Institute of Ceramics, Shanghai - P.R.China 200050}

\author{M.~Heikinheimo}
\affiliation{Helsinki Institute of Physics, Univ. of Helsinki, 00014 Helsinki - Finland}

\author{K.~Huitu}
\affiliation{Helsinki Institute of Physics, Univ. of Helsinki, 00014 Helsinki - Finland}

\author{M.~Kellermann}
\affiliation{Max-Planck-Institut f\"ur Physik, 80805 M\"unchen - Germany}

\author{R.~Maji}
\affiliation{Institut f\"ur Hochenergiephysik der \"Osterreichischen Akademie der Wissenschaften, 1050 Wien - Austria}
\affiliation{Atominstitut, Technische Universit\"at Wien, 1020 Wien - Austria}

\author{M.~Mancuso}
\affiliation{Max-Planck-Institut f\"ur Physik, 80805 M\"unchen - Germany}

\author{L.~Pagnanini}
\affiliation{Gran Sasso Science Institute, 67100 L'Aquila - Italy}
\affiliation{INFN - Laboratori Nazionali del Gran Sasso, 67010 Assergi - Italy}

\author{F.~Petricca}
\affiliation{Max-Planck-Institut f\"ur Physik, 80805 M\"unchen - Germany}

\author{S.~Pirro}
\affiliation{INFN - Laboratori Nazionali del Gran Sasso, 67010 Assergi - Italy}

\author{F.~Pr\"obst}
\affiliation{Max-Planck-Institut f\"ur Physik, 80805 M\"unchen - Germany}

\author{G.~Profeta}
\affiliation{Dipartimento di Scienze Fisiche e Chimiche, Universit\`a degli Studi dell'Aquila, 67100 L'Aquila - Italy}
\affiliation{INFN - Laboratori Nazionali del Gran Sasso, 67010 Assergi - Italy} 

\author{A.~Puiu}
\affiliation{Gran Sasso Science Institute, 67100 L'Aquila - Italy}
\affiliation{INFN - Laboratori Nazionali del Gran Sasso, 67010 Assergi - Italy}

\author{F.~Reindl}
\email{Corresponding author, florian.reindl@tuwien.ac.at}
\affiliation{Institut f\"ur Hochenergiephysik der \"Osterreichischen Akademie der Wissenschaften, 1050 Wien - Austria}
\affiliation{Atominstitut, Technische Universit\"at Wien, 1020 Wien - Austria}

\author{K.~Sch\"affner}
\email{Corresponding author, karoline.schaeffner@mpp.mpg.de}
\affiliation{Max-Planck-Institut f\"ur Physik, 80805 M\"unchen - Germany}

\author{J.~Schieck}
\affiliation{Institut f\"ur Hochenergiephysik der \"Osterreichischen Akademie der Wissenschaften, 1050 Wien - Austria}
\affiliation{Atominstitut, Technische Universit\"at Wien, 1020 Wien - Austria}

\author{D.~Schmiedmayer}
\affiliation{Institut f\"ur Hochenergiephysik der \"Osterreichischen Akademie der Wissenschaften, 1050 Wien - Austria}
\affiliation{Atominstitut, Technische Universit\"at Wien, 1020 Wien - Austria}

\author{C.~Schwertner}
\affiliation{Institut f\"ur Hochenergiephysik der \"Osterreichischen Akademie der Wissenschaften, 1050 Wien - Austria}
\affiliation{Atominstitut, Technische Universit\"at Wien, 1020 Wien - Austria}

\author{M.~Stahlberg}
\email{Corresponding author, martin.stahlberg@mpp.mpg.de}
\affiliation{Max-Planck-Institut f\"ur Physik, 80805 M\"unchen - Germany}

\author{A.~Stendahl}
\affiliation{Helsinki Institute of Physics, Univ. of Helsinki, 00014 Helsinki - Finland}

\author{F.~Wagner}
\affiliation{Institut f\"ur Hochenergiephysik der \"Osterreichischen Akademie der Wissenschaften, 1050 Wien - Austria}

\author{S.~Yue}
\affiliation{SICCAS - Shanghai Institute of Ceramics, Shanghai - P.R.China 200050}

\author{V.~Zema}
\email{Corresponding author, 
vanezema@mpp.mpg.de}
\affiliation{Max-Planck-Institut f\"ur Physik, 80805 M\"unchen - Germany}

\author{Y.~Zhu}
\affiliation{SICCAS - Shanghai Institute of Ceramics, Shanghai - P.R.China 200050}

\collaboration{The COSINUS Collaboration}
\author{A.~Bento}
\affiliation{Max-Planck-Institut f\"ur Physik, 80805 M\"unchen - Germany}
\affiliation{LIBPhys-UC, Physics Departments, University of Coimbra, 3004-516 Coimbra - Portugal}
\author{L.~Canonica}
\affiliation{Max-Planck-Institut f\"ur Physik, 80805 M\"unchen - Germany}
\author{A.~Garai}
\affiliation{Max-Planck-Institut f\"ur Physik, 80805 M\"unchen - Germany}

\date{\today}

\begin{abstract}
Low-temperature calorimeters based on a readout via Transition Edge Sensors (TESs) and operated below \unit[100]{mK} are well suited for rare event searches with outstanding resolution and low thresholds. We present first experimental results from two detector prototypes using a novel design of the thermometer coupling denoted \textit{remoTES}, which further extends the applicability of the TES technology by including a wider class of potential absorber materials. In particular, this design facilitates the use of materials whose physical and chemical properties, as e.g.~hygroscopicity, low hardness and low melting point, prevent the direct fabrication of the TES onto their surface. This is especially relevant in the context of the COSINUS experiment (Cryogenic Observatory for SIgnals seen in Next-Generation Underground Searches), where sodium iodide (NaI) is used as absorber material.
   
   With two remoTES prototype detectors operated in an above-ground R\&D facility, we achieve energy resolutions of \unit[$\sigma=87.8$]{eV} for a \unit[2.33]{g} silicon absorber and \unit[$\sigma=193.5$]{eV} for a \unit[2.27]{g} $\upalpha$-TeO$_{2}$ absorber, respectively. RemoTES calorimeters offer -- besides the wider choice of absorber materials -- a simpler production process combined with a higher reproducibility for large detector arrays and an enhanced radiopurity standard.
\end{abstract}

\maketitle

\section{Introduction}

Low-temperature detectors operated in the temperature region of \unit[10-100]{mK} make up an important detector class in particle physics. The detection of particles in cryogenic calorimeters is based on the measurement of the tiny temperature increase caused by an energy deposition in an absorber material,  by using very sensitive temperature sensors, e.g.~Transition Edge Sensors (TESs), Neutron Transmutation Doped-thermistors (NTDs), Kinetic Inductance Detectors (KIDs), or Metallic Magnetic Calorimeters (MMCs). 

A TES consists of a superconductive thin film operated in the state between the normal conducting and the superconducting phase. Due to the strong temperature dependence of the film resistance in the transition, even tiny temperature increases ($\mathcal{O}$(\textmu K)) induced by very small energy depositions ($\mathcal{O}$(KeV)) in the absorber, lead to  measurable changes in resistance of $\mathcal{O}$($10^{-1}\Omega$). For many decades, TESs have been a main pillar in the scope of cryogenic imaging spectrometers for earth and space applications; their use in the rare event search for direct dark matter detection~\cite{abdelhameed_first_2019,supercdms_collaboration_light_2021,supercdms_collaboration_search_2019,angloher_cosinus_2016} and CE$\upnu$NS~\cite{angloher_exploring_2019,billard_coherent_2017} by using massive absorbers $\mathcal{O}$(\unit[1-100]{g}) is a very active field of research at the frontier of knowledge.

TESs based on superconducting tungsten thin-films have been developed within the CRESST dark matter search at the Max-Planck Institute for Physics (MPP) in Munich over the last 30 years. The current stage is CRESST-III, which uses \unit[200]{nm}-thick tungsten films directly fabricated onto the surface of absorbers made of Al$_{2}$O$_{3}$, CaWO$_{4}$, LiAlO$_{2}$ and Si. The extraordinary high sensitivity achieved with TESs is one of the cornerstones of CRESST, which is the leading cryogenic experiment for low-mass dark matter search \cite{abdelhameed_first_2019, angloher_results_2017}.

The deposition of the TES thermometer directly onto the absorber ensures an excellent transmission of the non-thermal phonons to the W-film; it is the preferred solution as long as the absorber material can withstand \footnote{Nonetheless the fabrication process might induce stress in the crystal lattice, which is one of the main potential origins of the excess signal at low energies observed in several experiments (\cite{Proceedings:2022hmu})} the conditions during the involved fabrication steps, i.e. electron-beam evaporation, sputtering, chemical etching and photolithography. However, absorber materials that have a low melting point and/or are hygroscopic cannot undergo these processes.~A work-around solution for these delicate materials, such as for the~sodium iodide (NaI) crystals used in the COSINUS experiment \cite{angloher_cosinus_2016}, is the so-called composite design \cite{kiefer_composite_2009}: the TES is fabricated onto a separate substrate named ``carrier crystal" which in turn is connected to the absorber via an amorphous interface (glue, oil, or grease). This approach requires the phonons produced by a particle interacting in the absorber to propagate through the interface, the carrier crystal, and finally couple to the TES. A fraction of the signal is lost in this process due to the acoustic mismatch between the different materials of the carrier and the absorber~\cite{gray2012nonequilibrium}. In COSINUS we used Al$_{2}$O$_{3}$, Si, CaWO$_4$ and CdWO$_4$, which all have a high acoustic mismatch to the NaI absober.  Furthermore, in the case of a very good scintillator like NaI, a carrier crystal which is not fully transparent to the scintillation light will reabsorb part of the light, causing an additional power input to the TES. In this case, the resulting pulse shape in the TES will be a superposition of the NaI scintillation signal and the primary thermal signal~\cite{Zema:2020mkm}.

In \cite{pyle2015optimized}, the authors propose a TES-based detector concept, denoted \textit{remoTES} in the following, which does not suffer from these disadvantages, features a simpler production process, and additionally promises a better detector reproducibility. In a \textit{remoTES} detector, the TES is fabricated onto a separate wafer. The absorber crystal is then equipped with a gold pad that transmits the phonon signal created from an interaction in the absorber to the TES via a gold bonding wire. 
To our knowledge, experimental works using the TES coupling design proposed in~\cite{pyle2015optimized}  are absent to present date. However, the experiment RICOCHET is considering this design to read out superconducting absorbers for precision measurements of coherent elastic neutrino-nucleus scattering (CE$\upnu$NS)~\cite{Chen:2021tap} and the AMORE experiment~\cite{Alenkov:2019jis} utilizes thin gold films as phonon absorbers connected to a MMC~\cite{KIM2004208}.

Marrying ease of fabrication with a wide choice of absorber materials, the \textit{remoTES} may be a breakthrough for TES-instrumented calorimeters based on delicate absorber materials like NaI. In this article, we present the first experimental implementation and the first results which demonstrate the potential of such detectors.

\section{Detector design}
\subsection{Design of the \textit{remoTES}}
\begin{figure}[!htb]
\centering
   \includegraphics[width=0.5\textwidth]{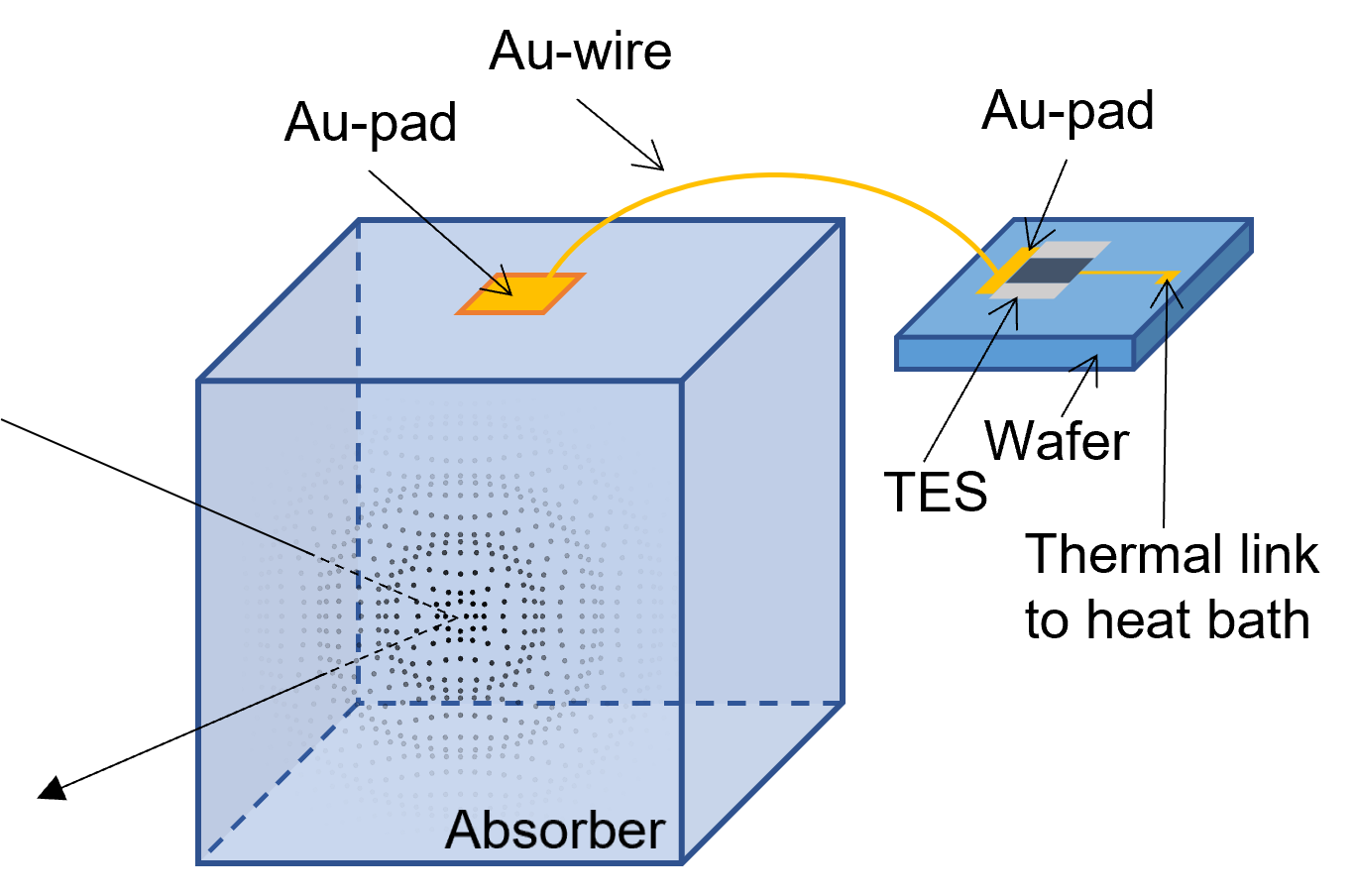}
  \caption{Schematic view of a \textit{remoTES} detector. The wafer where the TES is fabricated onto (right) is separated from the absorber crystal (left). This crystal is equipped with a gold pad which is linked directly to the TES via a gold bonding wire.}
  \label{fig:remoTES}
\end{figure}

The design of a \textit{remoTES} detector is schematically depicted in Figure~\ref{fig:remoTES}. It avoids the carrier crystal by coupling the TES directly to the target crystal via a gold pad connected to a gold bonding wire. A significant advantage of this design is that it eliminates the potential phonon barriers, i.e.~the loss of signal due to reflections caused by the acoustic mismatch between absorber crystal, interface and carrier crystal. Even more importantly, phonons directly couple to the electronic system of the gold pad via electron-phonon coupling~\cite{little1959transport}. In the literature, this coupling in gold is found to be comparable to or larger than in tungsten~\cite{probst_model_1995, sisti2001massive, karvonen2004electron,hart2009phase, pyle2015optimized}.~However, the surface and the thickness of the gold pad have to be carefully optimized: a larger surface enhances the phonon collection efficiency and hence the signal amplitude; at the same time a larger volume increases the total heat capacity, reducing the signal amplitude.

\subsection{Design of two \textit{remoTES}-prototypes}
In this manuscript we report the results from two different absorber materials, silicon (Si, \unit[2.33]{g}) and tellurium dioxide ($\upalpha$-TeO$_2$, \unit[2.27]{g}), read out by a \textit{remoTES}. These materials were selected to gain a fundamental understanding of the crucial parameters of this new TES-based readout design. We selected Si as it is a highly-studied standard material for low-temperature application and, thus, useful as a benchmark. We chose TeO$_{2}$ as a first step on the way to NaI, mainly for its similar solid-state properties, in particular regarding the heat capacity and the gap between the acoustic and optical phonon modes~\cite{jainmaterials, hinuma2017band}. 

\paragraph*{Technical details:} On the Si absorber (\unit[(20x10x5)]{mm$^3$}) a gold film of circular shape (\unit[3]{mm} diameter) of \unit[200]{nm} thickness was deposited by Magnetron-sputtering. The connection between the Au-pad on the Si absorber and the TES on the wafer was provided via a \unit[17]{$\upmu$m} thick and \unit[2-3]{mm} long Au-bonding wire. While the Au-wire was wedge-bonded onto the Au-pad of the TES, it was connected to the Au-pad on the Si absorber by using a silver-loaded epoxy (\unit[$\sim$25]{$\upmu$g}). Avoiding wedge-bonding on the Au-pad of the absorber opens up the possibility to also introduce very soft and fragile absorber crystals to this field of research, which would be damaged by solid phase welding processes (e.g.~wedge and ball bonding).
Alternatively, the Au-pad on the TeO$_{2}$ absorber (\unit[(20x10x2)]{mm$^3$}) consists of a \unit[400]{nm} thick Au-foil glued onto its surface by using a two-component epoxy resin. Using Au-foils which reveal high quality and increased thickness (Residual Resistance Ratio (RRR), listed in Table \ref{tab:summary_remoTES}), allow to exclude the absorber material from any kind of thin-film deposition processes which require high vacuum conditions. The electrical and thermal contact between Au-pad on the TeO$_{2}$ and the TES was realized with a wedge-bond on both sides, respectively. Table \ref{tab:summary_remoTES} summarizes the condensed information on both prototype detectors.

The TES on the thin Al$_{2}$O$_{3}$ wafer (\unit[(10x10x0.4)]{mm$^3$}) was identical for both measurements. It consists of a \unit[100]{nm}-thick tungsten film of an area of \unit[(220x300)]{\textmu m$^2$} and two aluminum pads for biasing of the sensor. The area of the W-film not overlapping with the pads is \unit[(70x300)]{$\upmu$m$^2$}. The weak thermal link to the heat bath consists of a \unit[1]{mm} long Au-stripe (\unit[20]{$\upmu$m} in width, \unit[80]{nm} in thickness). 

\begingroup
 \setlength{\tabcolsep}{8pt}
\renewcommand{\arraystretch}{1.2}
\begin{table}
   \centering

    \begin{tabular}{c || c  c}
        \multicolumn{3}{c}{}\\
 \textbf{Absorber material} & Si & TeO$_{2}$ \\
            \hline
                     \textbf{Absorber volume (mm$^3$)} & 20x10x5   & 20x10x2 \\
            \hline
\textbf{Au-pad properties} & 200 nm & 400 nm foil \\
                           & sputtered & glued \\
                           & RRR=3.79  & RRR=15  \\
            \hline
 \textbf{Au-wire properties}  & 17 \textmu m  & 17 \textmu m \\
                              & glued on pad & 2 wedge bonds\\
            \hline 
 \textbf{TES} & W-TES on Al$_{2}$O$_{3}$ & W-TES  on Al$_{2}$O$_{3}$ \\
            \hline
 \textbf{Energy resolution (eV)} & $87.8\pm5.6$  & $193.5\pm3.1$ \\[+2em]
    \end{tabular}
    \caption{Summary of the two measurements using the \textit{remoTES} design. RRR = Residual Resistance Ratio}
    \label{tab:summary_remoTES}
\end{table}
\endgroup

\section{Experimental results and discussion}

The measurements reported here were carried out in a dilution refrigerator of the CRESST group at the Max-Planck Institute for Physics in Munich. This is an above-ground, liquid $^4$He-precooled dilution refrigerator for R\&D measurements and TES testing. It is equipped with four SQUID channels (APS company) and a VME-based data acquisition using a hardware-triggered readout. The total exposure was \unit[1.06]{g$\cdot$d} for the Si detector and~\unit[2.28]{g$\cdot$d} for TeO$_2$. For means of energy calibration both prototype detectors were irradiated by an uncollimated $^{55}$Fe-source shining onto the bottom side of the absorbers. Furthermore, the TeO$_{2}$ detector was equipped with an additional collimated $^{55}$Fe-source irradiating solely the Au-pad glued onto the absorber. 

\subsection{Data analysis}
\label{subsec:dataanalysis}

\begin{figure}[!htb]
\centering
\begin{subfigure}[b]{0.49\textwidth}
   \includegraphics[width=\textwidth]{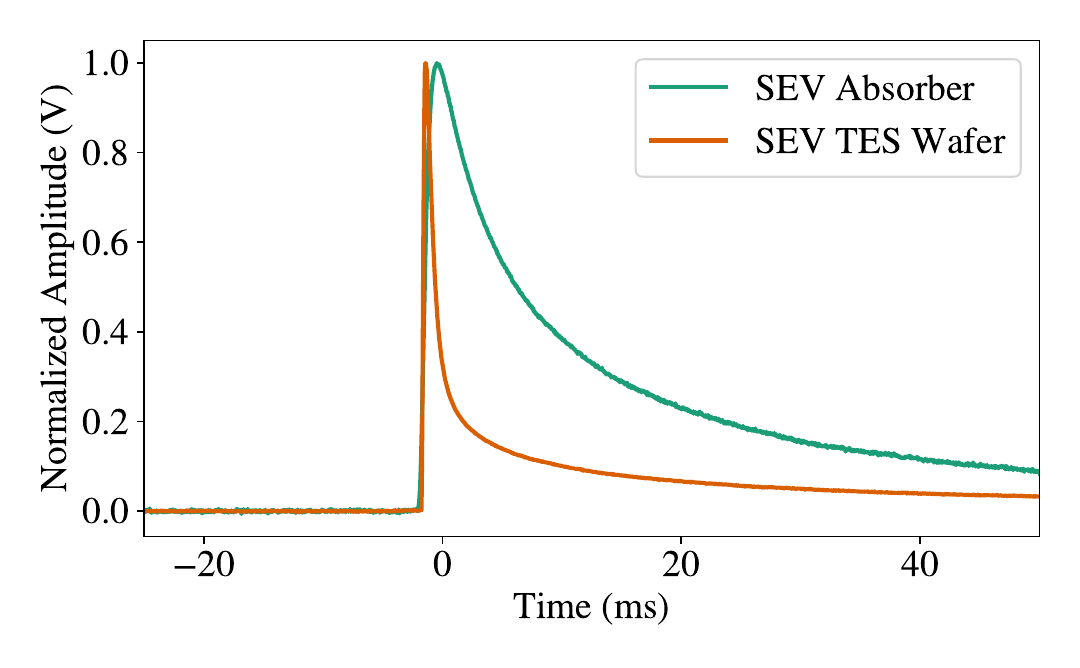}
   \caption{}
   \label{fig:sev566}
\end{subfigure}
\hfill
\begin{subfigure}[b]{0.49\textwidth}
   \includegraphics[width=\textwidth]{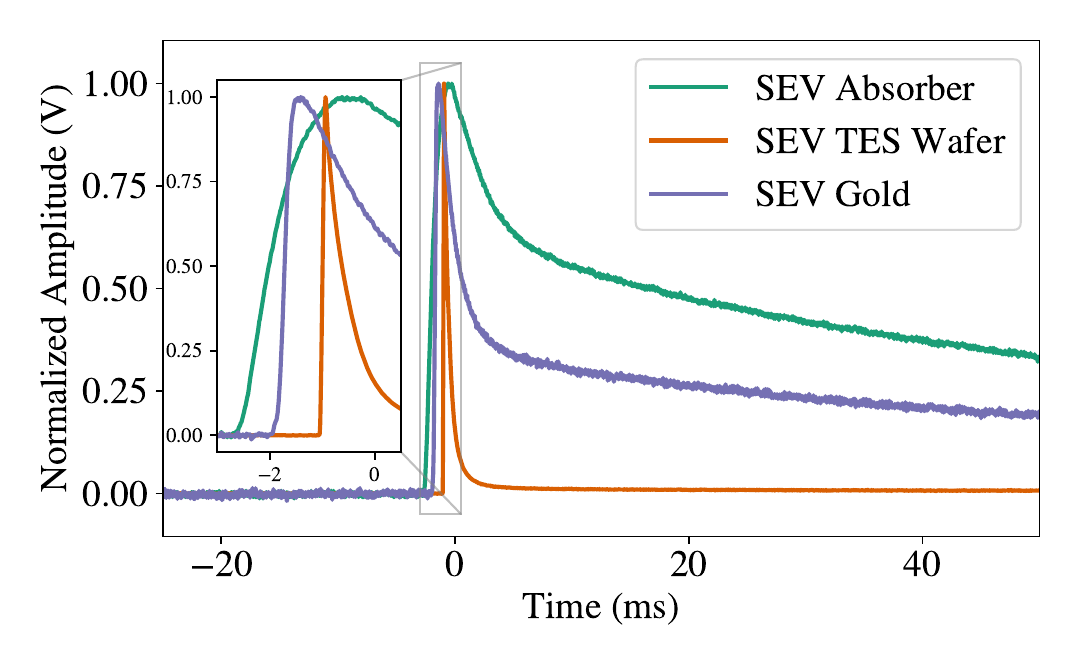}
   \caption{}
   \label{fig:sev573}
\end{subfigure}
  \caption{Standard events (SEVs) for the two prototype detectors. (a) Events in the Si absorber (green) and the TES wafer (orange). (b) Events in the TeO$_2$ absorber (green), the TES wafer (orange) and the gold pad on the TeO$_2$ (violet). The pulse onsets for each event class are slightly shifted for visualization purposes.}
  \label{fig:sev}
\end{figure}

We observe different event classes in both detectors. Standard events (SEVs $=$ averaged pulses scaled to a maximum amplitude of \unit[1]{V}) for these classes are shown in Figure \ref{fig:sev}; they were created from a small, selected sample of pulses with the same shape for each class. The most prevalent pulse shape, depicted in green, features a rise time of $\sim$\unit[1]{ms}, with a decay time of approximately \unit[44]{ms} for the Si prototype, and $\sim$\unit[122]{ms} for the TeO$_2$ prototype. The high rate of these events is consistent with hits in the absorber, considering its large mass in comparison with the mass of the wafer carrying the TES. Hits in the latter (orange in Figure \ref{fig:sev}) occur at a lower rate, and feature much faster signals: the rise times are only \unit[0.16]{ms} (Si measurement) and \unit[0.07]{ms} (TeO$_2$ measurement), while the decay times are $\sim$\unit[11.2]{ms} and $\sim$\unit[1.2]{ms}, respectively.

\begin{figure}[!htb]
\centering
\begin{subfigure}[b]{0.49\textwidth}
   \includegraphics[width=\textwidth]{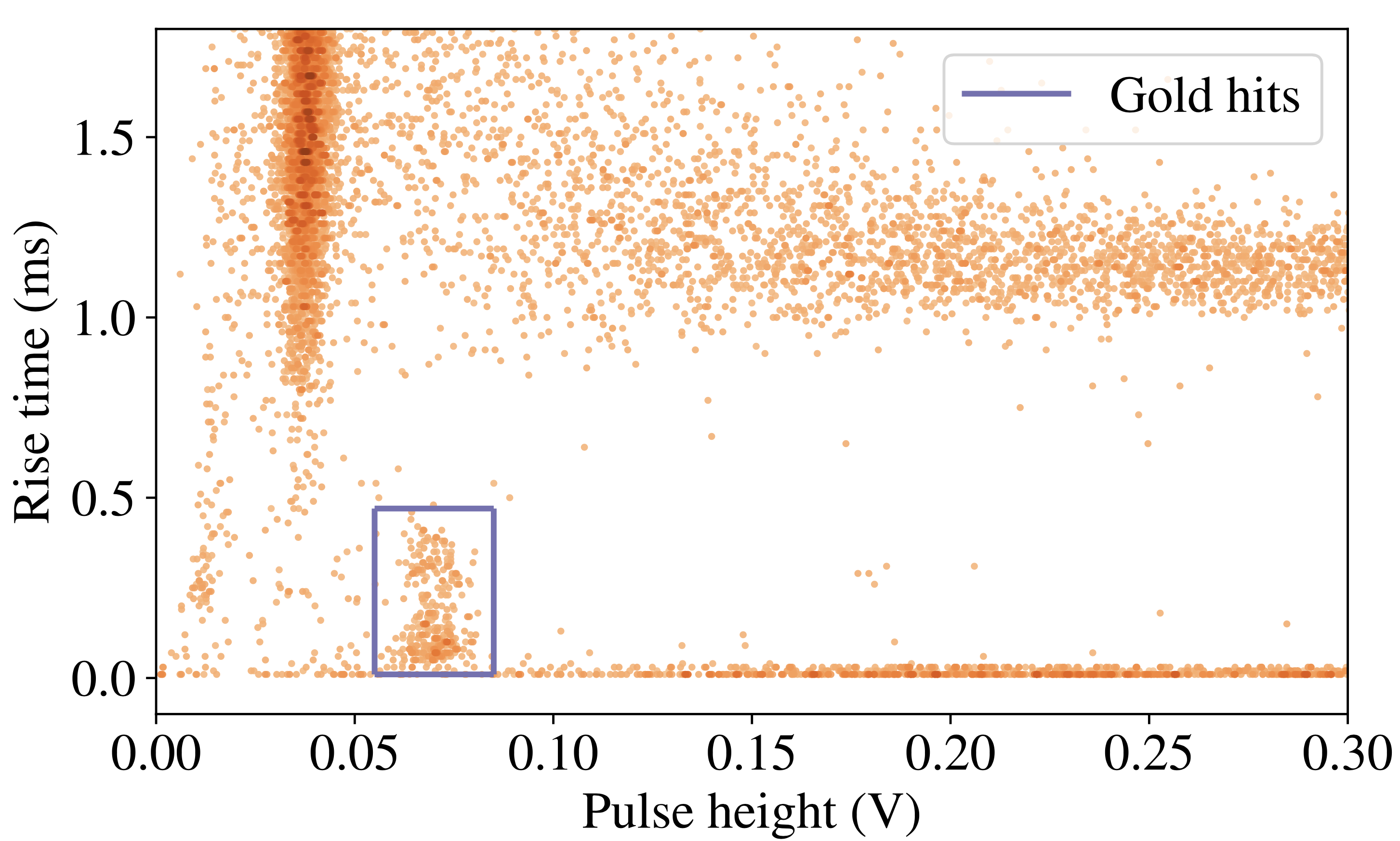}
   \caption{}
   \label{fig:gold_risetime}
\end{subfigure}
\hfill
\begin{subfigure}[b]{0.49\textwidth}
   \includegraphics[width=\textwidth]{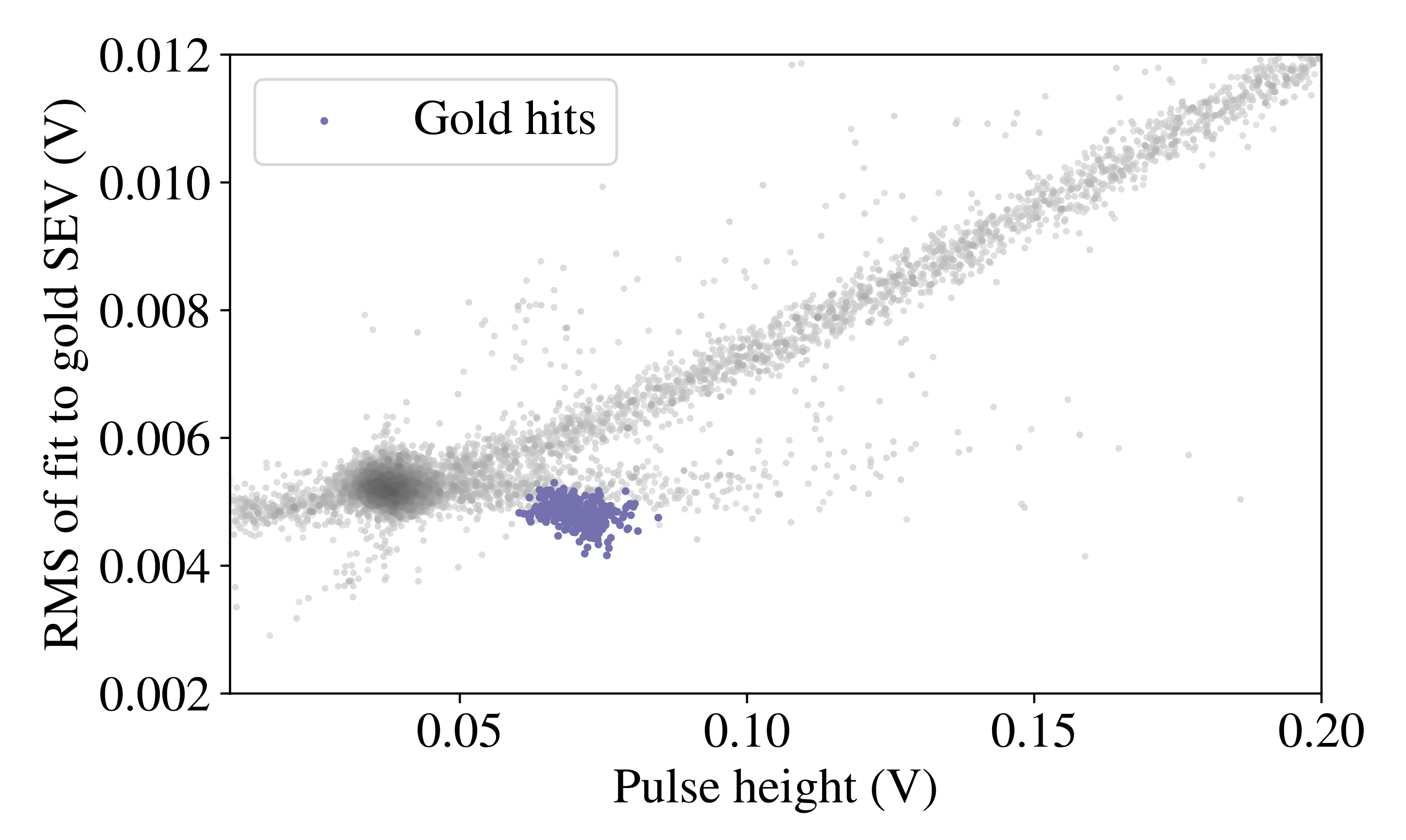}
   \caption{}
   \label{fig:fit_rms}
\end{subfigure}
  \caption{TeO$_2$-\textit{remoTES} data set: (a) Rise time versus moving average pulse height distribution. The violet box encloses the events in the gold foil produced by the collimated $^{55}$Fe-
 source. (b) Fit RMS for the gold SEV as a function of the pulse height distribution. The events from the violet box in  panel (a) are tagged and depicted in violet.}
  \label{fig:gold_rms}
\end{figure}

The TeO$_2$ prototype exhibits an additional event class. We identify these comparatively rare events (cf.~Table \ref{tab:event_rates}) as hits from the collimated $^{55}$Fe-source irradiating the Au-pad on the absorber. The rise time of the pulse is driven by the time needed to transport energy to the TES which depends on the energy carriers. The transport is fast for electrons and non-thermal phonons, which are the dominant energy carriers for the events created by interactions in the Au-pad or the carrier. In the TeO$_2$ absorber, a larger part of the initial phonon population undergoes thermalization, consequently leading to slower pulses. The Au-pad events are visible in Figure~\ref{fig:gold_risetime} as a cluster of events in the violet box. Their decay time is faster than the one of absorber events, but slower than the one of wafer events. The explanation can be two-fold: firstly, energy deposited in the Au-pad is partially transmitted back and forth to and from the absorber, leading to a larger decay time; secondly, the heat capacity of Au affects the decay time, leading to even slower pulses. As expected, the Au-pad events in Figure~\ref{fig:gold_risetime} are distributed around a larger pulse height ($\sim$~\unit[0.07]{V}) with respect to the pulse height of the $^{55}$Fe-events in the absorber ($\sim$~\unit[0.04]{V}). A definitive interpretation of the pulse shape requires further studies, which are currently ongoing.

\begin{figure}[!htb]
\centering
\includegraphics[width=0.8\textwidth]{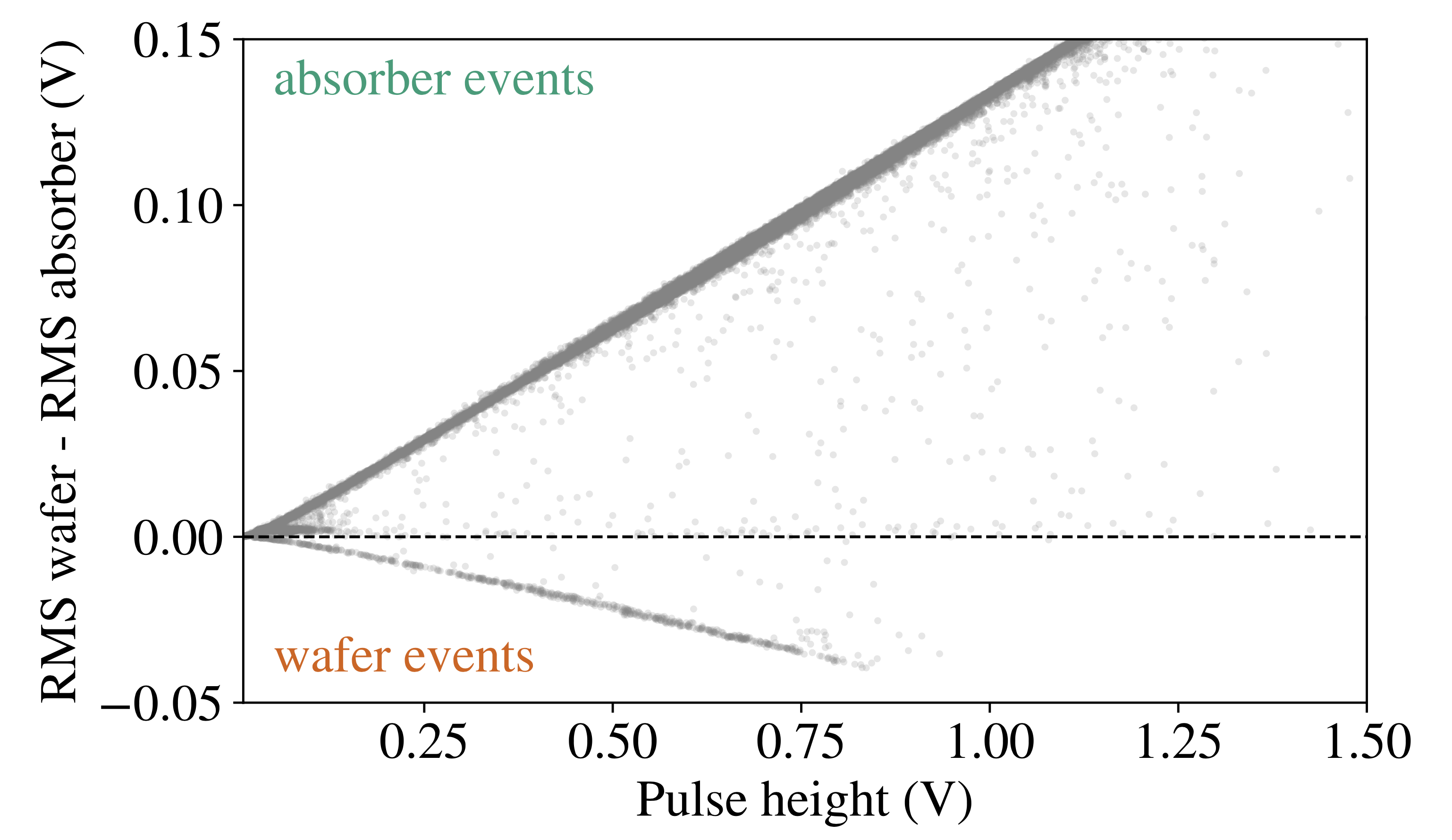}
\caption{Difference between RMS values from the wafer SEV fit and the absorber SEV fit as a function of moving-average pulse height for the TeO$_2$ prototype; different event bands corresponding to the different classes are visible. The event population between the absorber and the wafer band corresponds to the Au hits; its pulse shape is a mixture of the former two classes, and their shapes match it equally well.}
\label{fig:RMSDiff}
\end{figure}

Few basic quality cuts are applied to the complete dataset:  events coinciding with a decaying baseline from a previous pulse are removed by placing a cut on the pre-trigger baseline slope. Additionally, events with a reconstructed pulse onset more than \unit[1]{ms} away from the trigger position are discarded. To separate different event classes reliably, the root mean square (RMS) from a truncated standard event fit (see e.g. \cite{angloher_results_2017}) is used. We fit each pulse with all three SEVs, and the dataset for each event class contains only those events where the respective fit yields the lowest RMS. As an example, the discrimination between absorber and wafer events is shown in Figure \ref{fig:RMSDiff}. Figure \ref{fig:gold_rms} illustrates the discrimination between absorber events and Au-pad events, where the latter are taken from Figure \ref{fig:gold_risetime} and marked in purple. Table \ref{tab:event_rates} shows the number of surviving events for each prototype and event class; as stated before, absorber events are the dominant event class in both measurements, and Au-hits are less frequent than wafer hits for the TeO$_{2}$ detector.

\begingroup
\setlength{\tabcolsep}{8pt}
\renewcommand{\arraystretch}{1.2}
\begin{table}
   \centering
    \begin{tabular}{c c c c} 
        \multicolumn{4}{c}{}\\
 \textbf{Absorber} & \textbf{"Absorber"} & \textbf{"Wafer"}  & \textbf{"Gold"}\\
                     \textbf{material} 	& \textbf{events} & \textbf{events}  & \textbf{events} \\
                    \hline
                      Si                 & 6072            &  398      &  -   \\
                      TeO$_2$            & 28555            &  711      &  314   \\
            \hline
    \end{tabular}
    \caption{Number of surviving events after quality cuts for each measurement and event class (see text for details).}
    \label{tab:event_rates}
\end{table}
\endgroup

In addition to the events produced by particles interacting in the active detector parts, we periodically inject electrical pulses with known amplitudes into the heater of each detector. These heater pulses allow monitoring the detector response over time, and accounting for the effect of saturation in the TES, which influences the pulse shape at high energies (see e.g. \cite{angloher_results_2017}). A truncated fit with a dedicated heater SEV is performed for all heater pulses, and the resulting amplitude is used in combination with the input pulser setting to compensate non-linearities and time-variations of the reconstructed (fitted) amplitude for particle-induced events~\cite{angloher_limits_2002}. 

\subsection{Energy calibration and resolution}

\begin{figure}[!htb]
\centering
\begin{subfigure}[b]{0.49\textwidth}
   \includegraphics[width=\textwidth]{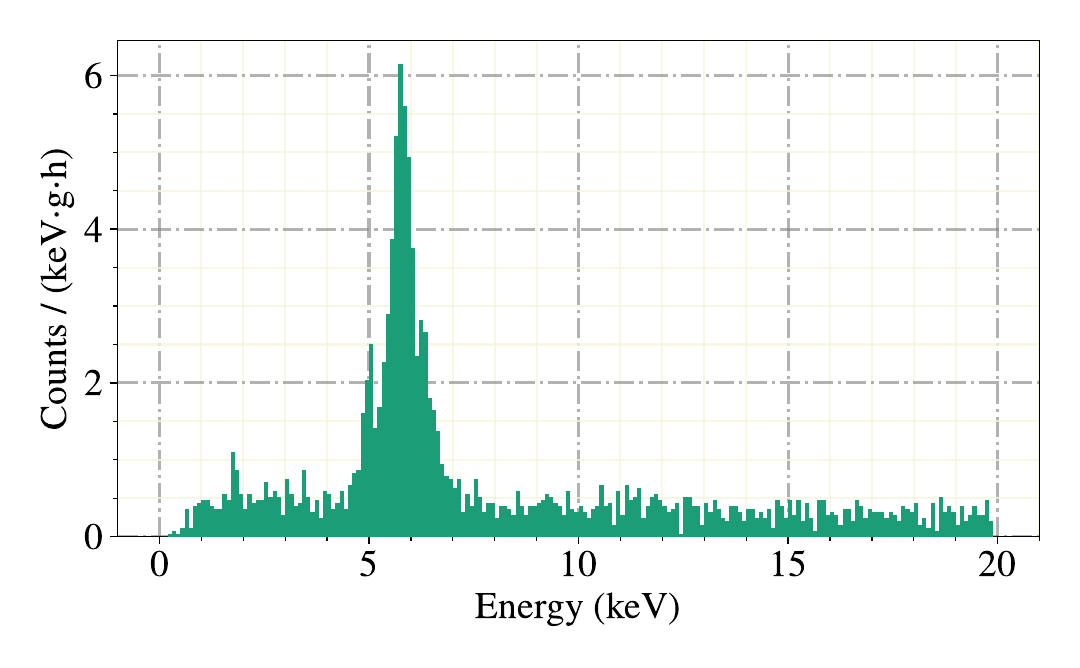}
   \caption{}
   \label{fig:spectrum566}
\end{subfigure}
\hfill
\begin{subfigure}[b]{0.49\textwidth}
   \includegraphics[width=\textwidth]{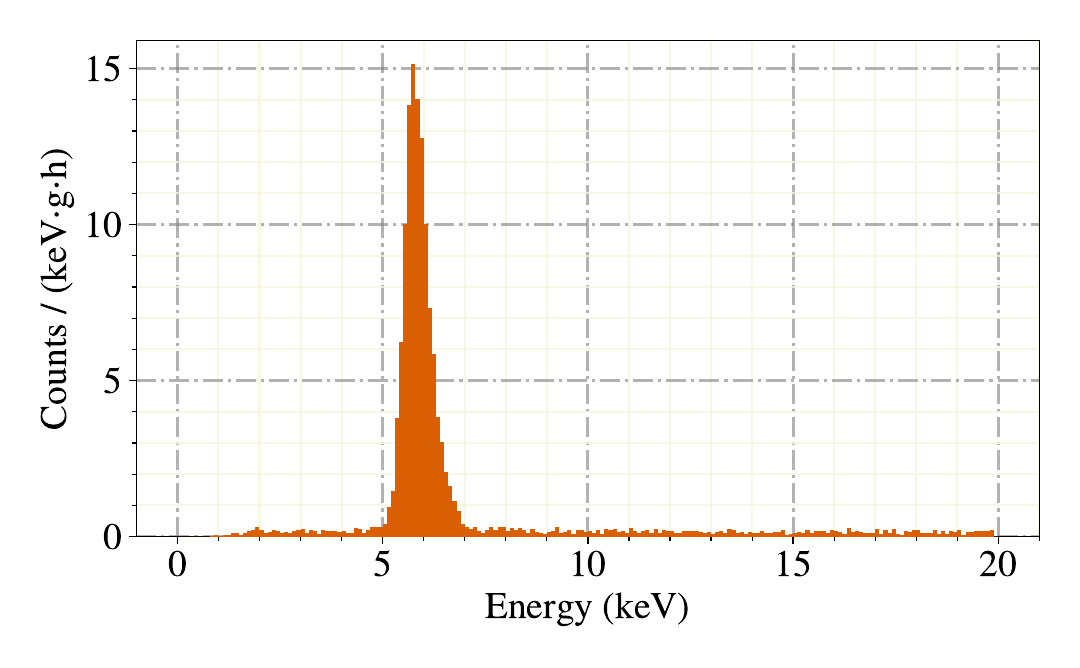}
   \caption{}
   \label{fig:spectrum573}
\end{subfigure}
  \caption{The energy spectra of the two prototype detectors: (a) Si absorber and (b) TeO$_2$ absorber. The intensity of the $^{55}$Fe-source producing X-rays of \unit[5.89]{keV} (K$_\alpha$) and \unit[6.49]{keV} (K$_\beta$) was significantly stronger for (b). The additional peaks in the Si detector ($\sim$\unit[1.8]{keV} and $\sim$\unit[5]{keV}) are consistent with x-ray emission from the K-shell of Si (\unit[1.84]{keV}), and an escape line due to the Cu holder (Cu L$_\alpha$ at \unit[0.93]{keV}).}
  \label{fig:spectrum}
\end{figure}

The K$_\alpha$ peak from the $^{55}$Fe source (\unit[5.89]{keV}) prominent in the spectrum of the Si and TeO$_2$ detector is used to calibrate the detector response, which is given by the truncated fit result for absorber events corrected by the heater response information. We chose the truncated fit over the usual method of optimal filtering \cite{gatti_processing_1986} here, since the noise conditions in both measurements changed over time, and the filter relies on a stable noise power spectrum. The resulting spectra for absorber events are given in Figure \ref{fig:spectrum}.   

The baseline resolution is determined for each detector in the following way: in each measurement, records without a trigger event are taken in regular intervals. These "empty baselines" are then cleaned, removing random coincidences with pulses and artifacts. Subsequently, we superimpose the SEV for a given detector with all remaining empty baselines. These artificial events (122 for the Si detector and 1932 for TeO$_2$) are passed through the analysis chain and reconstruction algorithm, and the detector resolution is extracted from the reconstructed amplitudes assuming that the noise fluctuations follow a Gaussian distribution. The resolutions (all in $\sigma$) are \unit[($18.4\pm1.2)$]{mV} for the Si and \unit[($1.22\pm 0.02)$]{mV} for the TeO$_2$ absorber, corresponding to \unit[($87.8\pm5.6$)]{eV} and  \unit[($193.5\pm3.1$)]{eV}, respectively (cf.~Table \ref{tab:summary_remoTES}). Typically, the achievable threshold of a TES is about five times the baseline resolution.

\section{Conclusion}

The \textit{remoTES} is an alternative design of the TES coupling to the absorber in a cryogenic calorimeter. It can be used in combination with delicate absorber materials such as NaI, which cannot withstand the process of direct TES fabrication. In this work we presented the successful operation of two \textit{remoTES} detector prototypes using a Si and a TeO$_2$ absorber in above-ground measurements, achieving baseline resolutions of \unit[87.8]{eV} and \unit[193.5]{eV}, respectively. We observed two event classes with distinctively different pulse shapes in each detector, which result from particle interactions in the absorber and in the wafer of the TES. Additionally, a third event class was visible in the TeO$_2$ measurement, which featured a very low rate and long decay time, consistent with hits in the gold pad glued onto the TeO$_2$ absorber. We already performed first measurements with NaI absorbers which are subject of future publications.
Compared to ``composite-type" detectors, the \textit{remoTES} improves phonon propagation from the absorber to the sensor, if absorber and carrier are made of different materials (\cite{schaffner2015particle}). This design additionally prevents the undesirable collection of scintillation light in the carrier. Furthermore, the separately fabricated wafer-TES can be reused with a different target -- this implies an improved reproducibility for \textit{remoTES} detectors with respect to their carrier design counterparts, and greatly facilitates comparisons between different absorber materials. In summary, our results show that the TES technology can be successfully used with almost any absorber material, and that \textit{remoTES} detectors are well suited for cryogenic rare event searches. Indeed, absorber materials that required the use of a carrier crystal in the past can now be operated without such a crystal by simply adding a gold interface. Even for ``standard" absorbers, the use of a \textit{remoTES} avoids exposing the material to fabrication processes, thus preserving the initial radiopurity and lattice condition of the absorber crystal.

\section*{Acknowledgments}
This work was carried out in the frame of the COSINUS collaboration. We would like to acknowledge Matt Pyle, Enectali Figueroa-Feliciano and Bernard Sadoulet who had the idea of developing such-kind of TES-based detectors, as published already in 2015 on arXiv:1503.01200 \cite{pyle2015optimized}. The data analysis was mainly carried out with collaboration-internal software tools, and partially with the open source Python package CAIT\cite{wagner_felix_2021_5091416}. We want to thank the MPP mechanical workshop team for in-time support in all needs regarding the cryogenic facility and detector holder fabrication. We are  grateful to the colleagues from the CRESST MPP group for sharing the infrastructure for detector production, the wet dilution refrigerator facility for the measurements and their cryogenic expertise and experience. In particular we thank Ahmed Abdelhameed for many insightful discussions on thin-film technology.

\bibliography{all,remoTES}

\begin{thebibliography}{26}
\providecommand{\natexlab}[1]{#1}
\providecommand{\url}[1]{\texttt{#1}}
\expandafter\ifx\csname urlstyle\endcsname\relax
  \providecommand{\doi}[1]{doi: #1}\else
  \providecommand{\doi}{doi: \begingroup \urlstyle{rm}\Url}\fi

\bibitem[Abdelhameed et~al.(2019)]{abdelhameed_first_2019}
A.~Abdelhameed et~al.
\newblock First results from the {CRESST}-{III} low-mass dark matter program.
\newblock \emph{Phys. Rev. D}, 100\penalty0 (10):\penalty0 102002, Nov. 2019.
\newblock \doi{10.1103/PhysRevD.100.102002}.

\bibitem[Alkhatib et~al.(2021)]{supercdms_collaboration_light_2021}
I.~Alkhatib et~al.
\newblock Light {Dark} {Matter} {Search} with a {High}-{Resolution} {Athermal}
  {Phonon} {Detector} {Operated} above {Ground}.
\newblock \emph{Phys. Rev. Lett.}, 127\penalty0 (6):\penalty0 061801, Aug.
  2021.
\newblock \doi{10.1103/PhysRevLett.127.061801}.

\bibitem[Agnese et~al.(2019)]{supercdms_collaboration_search_2019}
R.~Agnese et~al.
\newblock Search for low-mass dark matter with {CDMSlite} using a profile
  likelihood fit.
\newblock \emph{Phys. Rev. D}, 99\penalty0 (6):\penalty0 062001, Mar. 2019.
\newblock \doi{10.1103/PhysRevD.99.062001}.

\bibitem[Angloher et~al.(2016)]{angloher_cosinus_2016}
G.~Angloher et~al.
\newblock The {COSINUS} project: perspectives of a {NaI} scintillating
  calorimeter for dark matter search.
\newblock \emph{Eur. Phys. J. C}, 76\penalty0 (8):\penalty0 441, Aug. 2016.
\newblock ISSN 1434-6044, 1434-6052.
\newblock \doi{10.1140/epjc/s10052-016-4278-3}.

\bibitem[Angloher et~al.(2019)]{angloher_exploring_2019}
G.~Angloher et~al.
\newblock Exploring $\hbox {{CE}}\nu \hbox {{NS}}$ with {NUCLEUS} at the
  {Chooz} nuclear power plant.
\newblock \emph{Eur. Phys. J. C}, 79\penalty0 (12):\penalty0 1018, Dec. 2019.
\newblock ISSN 1434-6052.
\newblock \doi{10.1140/epjc/s10052-019-7454-4}.

\bibitem[Billard et~al.(2017)]{billard_coherent_2017}
J.~Billard et~al.
\newblock Coherent neutrino scattering with low temperature bolometers at
  {Chooz} reactor complex.
\newblock \emph{J. Phys. G: Nucl. Part. Phys.}, 44\penalty0 (10):\penalty0
  105101, Aug. 2017.
\newblock ISSN 0954-3899.
\newblock \doi{10.1088/1361-6471/aa83d0}.

\bibitem[Angloher et~al.(2017)]{angloher_results_2017}
G.~Angloher et~al.
\newblock Results on {MeV}-scale dark matter from a gram-scale cryogenic
  calorimeter operated above ground.
\newblock \emph{Eur. Phys. J. C}, 77\penalty0 (9):\penalty0 637, Sept. 2017.
\newblock ISSN 1434-6044, 1434-6052.
\newblock \doi{10.1140/epjc/s10052-017-5223-9}.

\bibitem[Adari et~al.(2022)]{Proceedings:2022hmu}
P.~Adari et~al.
\newblock {EXCESS workshop: Descriptions of rising low-energy spectra}.
\newblock \emph{arXiv:2202.05097}, 2 2022.
\newblock \doi{10.48550/arXiv.2202.05097}.

\bibitem[Angloher et~al.(2009)]{kiefer_composite_2009}
G.~Angloher et~al.
\newblock Composite {CaWO}$_4$ {Detectors} for the {CRESST}‐{II}
  {Experiment}.
\newblock In \emph{{AIP} {Conference} {Proceedings}}, volume 1185, pages
  651--654. AIP Publishing, Dec. 2009.
\newblock \doi{10.1063/1.3292426}.

\bibitem[Gray(2012)]{gray2012nonequilibrium}
K.~E. Gray.
\newblock Nonequilibrium superconductivity, phonons, and kapitza boundaries.
\newblock \emph{Springer Science \& Business Media}, 65, 2012.
\newblock \doi{10.1007/978-1-4684-3935-9}.

\bibitem[Zema(2020)]{Zema:2020mkm}
V.~Zema.
\newblock \emph{\emph{Unveiling the nature of dark matter with direct detection
  experiments}}.
\newblock PhD thesis, Gran Sasso Science Institute, Chalmers U. Tech., 2020.
\newblock URL \url{https://inspirehep.net/literature/1813204}.

\bibitem[Pyle et~al.(2015)]{pyle2015optimized}
M.~Pyle et~al.
\newblock Optimized designs for very low temperature massive calorimeters.
\newblock \emph{arXiv:1503.01200}, 2015.
\newblock \doi{10.48550/arXiv.1503.01200}.

\bibitem[Chen(2021)]{Chen:2021tap}
R.~Chen.
\newblock {Transition Edge Sensor Chip Design of Modular CE\ensuremath{\nu}NS
  Detector for the Ricochet Experiment}.
\newblock \emph{arXiv}, 11 2021.
\newblock \doi{10.48550/arXiv.2111.05757}.

\bibitem[Alenkov et~al.(2019)]{Alenkov:2019jis}
V.~Alenkov et~al.
\newblock {First Results from the AMoRE-Pilot neutrinoless double beta decay
  experiment}.
\newblock \emph{Eur. Phys. J. C}, 79\penalty0 (9):\penalty0 791, 2019.
\newblock \doi{10.1140/epjc/s10052-019-7279-1}.

\bibitem[Kim et~al.(2004)]{KIM2004208}
Y.~H. Kim et~al.
\newblock Measurements and modeling of the thermal properties of a calorimeter
  having a sapphire absorber.
\newblock \emph{Nuclear Instruments and Methods in Physics Research Section A:
  Accelerators, Spectrometers, Detectors and Associated Equipment},
  520\penalty0 (1):\penalty0 208--211, 2004.
\newblock ISSN 0168-9002.
\newblock \doi{10.1016/j.nima.2003.11.230}.
\newblock Proceedings of the 10th International Workshop on Low Temperature
  Detectors.

\bibitem[Little(1959)]{little1959transport}
W.~Little.
\newblock The transport of heat between dissimilar solids at low temperatures.
\newblock \emph{Canadian Journal of Physics}, 37\penalty0 (3):\penalty0
  334--349, 1959.
\newblock \doi{10.1139/p59-037}.

\bibitem[Pröbst et~al.(1995)]{probst_model_1995}
F.~Pröbst et~al.
\newblock Model for cryogenic particle detectors with superconducting phase
  transition thermometers.
\newblock \emph{Journal of Low Temperature Physics}, 100\penalty0
  (1-2):\penalty0 69--104, July 1995.
\newblock ISSN 0022-2291, 1573-7357.
\newblock \doi{10.1007/BF00753837}.

\bibitem[Sisti et~al.(2001)]{sisti2001massive}
M.~Sisti et~al.
\newblock Massive cryogenic particle detectors with low energy threshold.
\newblock \emph{Nuclear Instruments and Methods in Physics Research Section A:
  Accelerators, Spectrometers, Detectors and Associated Equipment},
  466\penalty0 (3):\penalty0 499--508, 2001.
\newblock \doi{10.1016/S0168-9002(01)00801-4}.

\bibitem[Karvonen et~al.(2004)]{karvonen2004electron}
J.~Karvonen et~al.
\newblock Electron--phonon interaction in thin copper and gold films.
\newblock \emph{physica status solidi (c)}, 1\penalty0 (11):\penalty0
  2799--2802, 2004.
\newblock \doi{10.1002/pssc.200405326}.

\bibitem[Hart et~al.(2009)]{hart2009phase}
S.~Hart et~al.
\newblock Phase separation in tungsten transition edge sensors.
\newblock In \emph{AIP Conference Proceedings}, volume 1185, pages 215--218.
  American Institute of Physics, 2009.
\newblock \doi{10.1063/1.3292318}.

\bibitem[Jain et~al.(2013)]{jainmaterials}
A.~Jain et~al.
\newblock The materials project: A materials genome approach to accelerating
  materials innovation.
\newblock \emph{APL Materials}, 1\penalty0 (011002), 2013.
\newblock \doi{10.1063/1.4812323}.

\bibitem[Hinuma et~al.(2017)]{hinuma2017band}
Y.~Hinuma et~al.
\newblock Band structure diagram paths based on crystallography.
\newblock \emph{Computational Materials Science}, 128:\penalty0 140--184, 2017.
\newblock \doi{10.1016/j.commatsci.2016.10.015}.

\bibitem[Angloher et~al.(2002)]{angloher_limits_2002}
G.~Angloher et~al.
\newblock Limits on {WIMP} dark matter using sapphire cryogenic detectors.
\newblock \emph{Astropart. Phys.}, 18\penalty0 (1):\penalty0 43--55, Aug. 2002.
\newblock ISSN 0927-6505.
\newblock \doi{10.1016/S0927-6505(02)00111-1}.

\bibitem[Gatti and Manfredi(1986)]{gatti_processing_1986}
E.~Gatti and P.~F. Manfredi.
\newblock Processing the signals from solid-state detectors in
  elementary-particle physics.
\newblock \emph{Riv. Nuovo Cim.}, 9\penalty0 (1):\penalty0 1--146, Jan. 1986.
\newblock ISSN 1826-9850.
\newblock \doi{10.1007/BF02822156}.

\bibitem[Sch{\"a}ffner et~al.(2015)]{schaffner2015particle}
K.~Sch{\"a}ffner et~al.
\newblock Particle discrimination in teo2 bolometers using light detectors read
  out by transition edge sensors.
\newblock \emph{Astroparticle Physics}, 69:\penalty0 30--36, 2015.
\newblock \doi{10.1016/j.astropartphys.2015.03.008}.

\bibitem[Wagner et~al.(2021)]{wagner_felix_2021_5091416}
F.~Wagner et~al.
\newblock fewagner/cait: v1.0.0, July 2021.
\newblock URL \url{https://doi.org/10.5281/zenodo.5091416}.

\end{thebibliography}

\end{document}